\documentclass[graybox,vecphys]{svmult}

\usepackage{type1cm}        
\usepackage{makeidx}         
\usepackage{graphicx}        
\usepackage{multicol}        
\usepackage[bottom]{footmisc}

\usepackage{newtxtext}       %
\usepackage{newtxmath}       
\usepackage{subfigure}

\usepackage{bm}

\makeindex             

\newcommand{\kn}{\mathrm{Kn}}
\newcommand{\DS}{\mathrm{dS}}
\newcommand{\DV}{\mathrm{d}\bm{x}}

\newcommand{\rmU}{\mathrm{U}}
\newcommand{\rmS}{\mathrm{S}}
\newcommand{\rmUU}{\mathrm{UU}}
\newcommand{\rmSS}{\mathrm{SS}}
\newcommand{\rmUS}{\mathrm{US}}
\newcommand{\rmI}{\mathrm{I}}
\newcommand{\rmII}{\mathrm{II}}
\newcommand{\rmw}{\mathrm{w}}
\newcommand{\rmH}{\mathrm{H}}
\newcommand{\Op}{O\!p}
\newcommand{\re}{\mathrm{Re}}
\renewcommand{\ij}{i\!j}
\newcommand{\J}{\mathrm{J}}

\allowdisplaybreaks[1]

\begin{document}

\title*{Inverse Magnus effect in a rarefied gas}
\titlerunning{Inverse Magnus effect}
\author{Satoshi Taguchi and Tetsuro Tsuji}
\institute{Satoshi Taguchi \at Department of Advanced Mathematical Sciences, Graduate School of Informatics, Kyoto University, Kyoto 606-8501, Japan. \email{taguchi.satoshi.5a@kyoto-u.ac.jp}
\and Tetsuro Tsuji \at Department of Advanced Mathematical Sciences, Graduate School of Informatics, Kyoto University, Kyoto 606-8501, Japan. \email{tsuji.tetsuro.7x@kyoto-u.ac.jp}}
%
%
\maketitle

\abstract*{The transverse force exerted on a rotating sphere immersed in an otherwise uniform flow of a rarefied gas is investigated based on the Bhatnagar--Gross--Krook (BGK) model of the Boltzmann equation assuming the Maxwell boundary condition on the sphere. In several existing studies, it has been shown that the transverse force acting on the sphere, also known as the Magnus force, has opposite signs in the free molecular and continuum flows. The present study intends to clarify the force's transition in terms of the Knudsen number (i.e., the reciprocal ratio of the sphere radius to the molecular mean free path) with a particular interest in the impact of the sphere's surface accommodation. It is found that the threshold of the Knudsen number, at which the transverse force changes the sign, depends only weakly on the accommodation coefficient, suggesting certain robustness in the threshold. The present study is an extension of the previous work [S. Taguchi and T. Tsuji, J. Fluid. Mech. 933, A37 (2022)], in which the case of complete accommodation (diffuse reflection) is exclusively considered.}

\abstract{The transverse force exerted on a rotating sphere immersed in an otherwise uniform flow of a rarefied gas is investigated based on the Bhatnagar--Gross--Krook (BGK) model of the Boltzmann equation assuming the Maxwell boundary condition on the sphere. In several existing studies, it has been shown that the transverse force acting on the sphere, also known as the Magnus force, has opposite signs in the free molecular and continuum flows. The present study intends to clarify the force's transition in terms of the Knudsen number (i.e., the reciprocal ratio of the sphere radius to the molecular mean free path) with a particular interest in the impact of the sphere's surface accommodation. It is found that the threshold of the Knudsen number, at which the transverse force changes the sign, depends only weakly on the accommodation coefficient, suggesting certain robustness in the threshold. The present study is an extension of the previous work [S. Taguchi and T. Tsuji, J. Fluid. Mech. 933, A37 (2022)], in which the case of complete accommodation (diffuse reflection) is exclusively considered.}


\section{\label{sec:intro}Introduction}
Understanding the motion of tiny particles in a gas is vital in many areas including micro-nano technology, control of particles, and aerosol transport. For this purpose, studying a gas flow around a particle is crucial. It is known that the Navier--Stokes equation ceases to be accurate in describing flows with the miniaturization of a particle. Alternatively, the Boltzmann equation, describing molecules' collective behavior, can explain flows around tiny objects \cite{Cercignani06,Sone07}. In this context, the gas is referred to as a rarefied gas.

In the present paper, we focus on the force, mainly the transverse (or lift) force, acting on a single rigid spherical particle moving in a rarefied gas. This transverse force occurs when rotation is given to the particle and is known as the Magnus effect. Interestingly, the transverse force depends significantly on the relative particle size to the mean free path of the gas molecules. Indeed, the authors showed, for the case of a small relative particle velocity, that the transverse force $\vec{F}_L$ is given by $\vec{F}_L = \pi \rho a^3 (\vec{\Omega} \times \vec{v}) \bar{h}_L$, where $\rho$ is the density of the surrounding gas, $a$ the radius of the sphere, $\vec{\Omega}$ the particle's angular velocity, and $\vec{v}$ the particle's velocity, and that the numerical factor $\bar{h}_L$ decreases monotonically from positive to negative values with the increase of the Knudsen number (i.e., the ratio between the mean free path of the gas molecules and the sphere radius) \cite{taguchi_tsuji_2022_jfm}. To the authors' best knowledge, this was the first theoretical result to show the force's transition to the so-called inverse Magnus effect, formerly known for a free molecular gas \cite{Wang_AIAA_1972,Ivanov-Yanshin_FD80,Borg-Soederholm-Essen-PHF03} and having drawn researchers' attention \cite{Weidman_Herczynski_PhysFluids_2004,Liu-Bogy_PoF_2008,Liu_Bogy_PhysFluids_2009,Volkov-JFM-2011,Baier+Tiwari+Shrestha+Klar+Hardt_PhysRevFluids_2018,Wang_Yu_Luo_Xia_Zong_PoF_2018,Shi_Rzehak_ChemEngSci_2019,Kumar_Dhiman_Reddy_PhysRevE_2019}. 

The present paper serves a dual purpose. First, we want to outline the analysis of \cite{taguchi_tsuji_2022_jfm}, which contains many technical details and formulas. The second purpose is to get insight on the impact of the sphere's surface accommodation into the transverse force.
In particular, we try to give the first answer to the following question: does the threshold of the Knudsen number for the negative lift depend much on the surface property of the sphere? To simplify the analysis as much as possible, we assume the Maxwell boundary condition on the sphere with accommodation coefficient $\alpha$ ($0< \alpha \le 1$).
As a result, we find that the threshold is insensitive to $\alpha$, although the transverse force is generally affected by $\alpha$. This robustness of the threshold may open a way to use the transverse force to separate small particles from large ones.

\begin{figure}[t]
    \centering
    \includegraphics[width=0.65\textwidth]{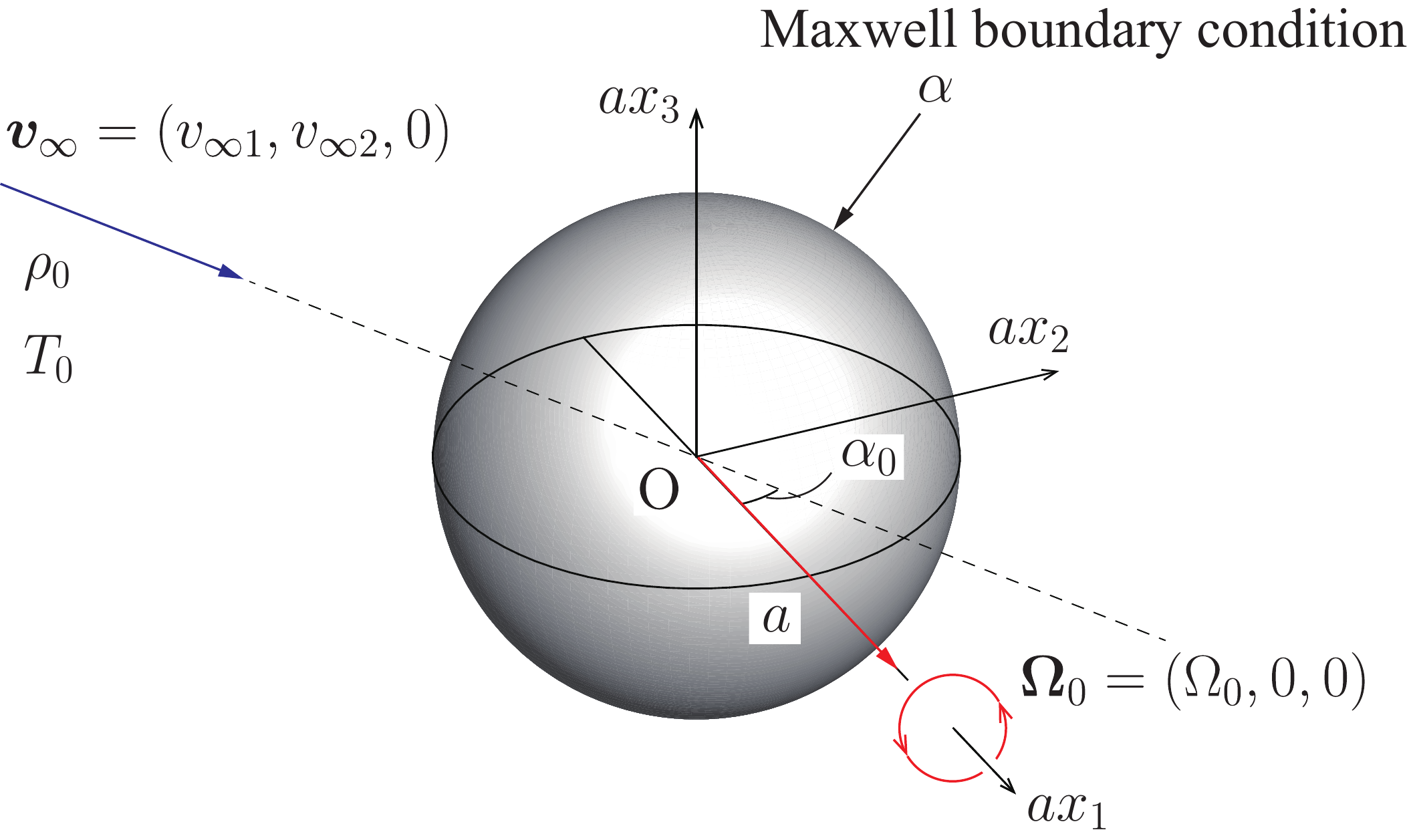}
    \caption{Schematic of the problem: a uniform flow past a rotating sphere. The Maxwell condition is assumed on the sphere surface with accommodation coefficient $\alpha$ ($0 < \alpha \le 1$).}
    \label{fig:problem}
\end{figure}

\section{\label{sec:problem_and_formulation}Problem and formulation}
In this section, we give the complete statement of the problem, followed by its mathematical formulation.

\subsection{\label{subsec:problem}Problem}
We consider a rigid sphere with radius $a$ placed in an otherwise uniform flow of a rarefied gas (see Fig.~\ref{fig:problem}). Let $a x_i$, $i=1,2,3$, (or $a \vec{x}$) denote the Cartesian coordinate system for the physical space, whose origin is at the sphere center. We assume that the state of the gas at infinity is a uniform flow, whose macroscopic velocity, density, and temperature are denoted by $\bm{v}_\infty=(v_{\infty 1},v_{\infty 2},0)$, $\rho_0$, and $T_0$, respectively.
We further suppose that the sphere is kept at a uniform temperature $T_0$ and
is rotating about one of its diameters with constant angular velocity $\bm{\Omega}_0 = (\Omega_0,0,0)$. External forces (such as gravity) are assumed to be absent. We investigate the steady behavior of the gas around the sphere under the following basic assumptions:

\begin{enumerate}
    \item The behavior of the gas is described by the Bhatnagar--Gross--Krook (BGK) model \cite{Bhatnagar-Gross-Krook54,Welander54} of the Boltzmann equation. In the sequel, however, we use the Boltzmann equation for a general analysis. The BGK model will be used only for the actual numerical computations.
    \item The gas molecules are reflected on the sphere according to Maxwell's diffuse-specular reflection condition.
    \item The macroscopic flow velocity at infinity is much smaller than the corresponding thermal speed, i.e., $|\bm{v}_\infty| \ll c_0$, where $c_0 = \sqrt{2RT_0}$. Here, $R$ is the specific gas constant, i.e., $R=k_B/m$ with $k_B$ and $m$ being the Boltzmann constant and the mass of a molecule, respectively.
    \item The sphere's equatorial velocity due to the rotation is of the same order as the flow speed at infinity, i.e, $a \Omega_0 /|\bm{v}_\infty| = O(1)$.
\end{enumerate}
For convenience, we also use $p_0 = \rho_0 RT_0$ to denote the gas pressure at the reference state.

\subsection{Formulation}
In this subsection, we formulate the problem mentioned in Set.~\ref{subsec:problem}. 
The formulation will be presented mainly for the Boltzmann equation. The reader is referred to \cite{taguchi_tsuji_2022_jfm} for the formulation based on the BGK model.

\subsubsection{Basic equations}

Let $c_0 \zeta_i$ (or $c_0 \vec{\zeta}$) denote the molecular velocity, and let $\rho_0 c_0^{-3} (1 + \phi(\vec{x},\vec{\zeta})) E$ the velocity distribution function of the gas molecules, where $E = \pi^{-3/2} \exp(-|\vec{\zeta}|^2)$.
The stationary Boltzmann equation for $\phi=\phi(\vec{x},\vec{\zeta})$ in the absence of external forces is reduced to \cite{Sone07}
\begin{align} \label{e:boltzmann} 
    \vec{\zeta}\cdot \nabla_{\vec{x}} \phi = \frac{1}{k} (\mathcal{L}(\phi) + \mathcal{J}(\phi,\phi)).
\end{align}
Here, $\mathcal{L}(\phi)$ is the linearized collision integral and
$\mathcal{J}(\phi,\phi)$ is the collision integral defined by
\begin{subequations}
\begin{align}
& \mathcal{J}(F,G) = \frac{1}{2} \int_{(\vec{\zeta}_*,\vec{e}) \in \mathbb{R}^3 \times \mathbb{S}^2} E_* (F'_* G' + F' G'_* - F_* G - F G_*) \,B \,\D\Omega(\vec{e}) \D\vec{\zeta}_*,
\label{e:collision}
\\
& F = F(\vec{\zeta}), \quad F_* = F(\vec{\zeta}_*), \quad F' = F(\vec{\zeta}'), \quad F_*' = F(\vec{\zeta}_*'),  
\\
& G = G(\vec{\zeta}), \quad G_* = G(\vec{\zeta}_*), \quad G' = G(\vec{\zeta}'), \quad G_*' = G(\vec{\zeta}_*'),  
\\
& \vec{\zeta}' = \vec{\zeta} + [(\vec{\zeta}_* - \vec{\zeta}) \cdot \vec{e}] \vec{e},
\quad \vec{\zeta}_*' = \vec{\zeta}_* - [(\vec{\zeta}_* - \vec{\zeta}) \cdot \vec{e}] \vec{e},
\end{align}
\end{subequations}
where $E_* = \pi^{-3/2} \exp(-|\vec{\zeta}_*|^2)$, $\vec{e}$ denotes the unit vector, $\D\Omega(\vec{e})$ the solid angle element in the direction of $\vec{e}$, and
$B=B\left(\frac{|\vec{e}\cdot(\vec{\zeta}_* - \vec{\zeta})|}{|\vec{\zeta}_* - \vec{\zeta}|},|\vec{\zeta}_*-\vec{\zeta}|\right)$ the non-negative function 
determined by a specific intermolecular force \cite{Sone07}. For example, $B = \frac{1}{4\sqrt{2\pi}}|\vec{e}\cdot (\vec{\zeta}_* - \vec{\zeta})|$ for a hard-sphere gas.
The integration on the r.h.s of \eqref{e:collision} is carried out for the whole space of $\vec{\zeta}_*$ and all directions of $\vec{e}$.
With $\mathcal{J}$ defined above, the linearized collision integral $\mathcal{L}(\phi)$ is given by
$\mathcal{L}(\phi) = 2 \mathcal{J}(1,\phi)$.
The parameter $k$ is defined by 
$k = \frac{\sqrt{\pi}}{2}\kn$, where $\kn = \ell_0/a$ is the Knudsen number with $\ell_0$ the molecular mean free path in the reference state at rest with density $\rho_0$ and temperature $T_0$. For a hard-sphere gas, $\ell_0=[\sqrt{2} \pi d_m^2(\rho_0/m)]^{-1}$ with $d_m$ and $m$ being the diameter and mass of a molecule, respectively.
For the BGK model of the collision integral, whose explicit form is omitted here \cite{taguchi_tsuji_2022_jfm}, $\ell_0 = \frac{2}{\sqrt{\pi}} \frac{c_0}{A_c \rho_0}$ with $A_c$ being a constant.

Let us introduce the spherical coordinate system $(r,\theta,\varphi)$ associated with $(x_1,x_2,x_3)$ by
$x_1 = r \cos \theta$, $x_2 = r \sin \theta \cos \varphi$, 
$x_3 = r \sin \theta \sin \varphi$ ($0 \le \theta \le \pi$, $0 \le \varphi < 2 \pi$). Henceforth, the components of vectors or tensors in the spherical coordinates are designated by attaching the subscript $r,\theta,\varphi$ (e.g., $\zeta_r$, $\zeta_\theta$, $\zeta_\varphi$).
The Maxwell boundary condition on the sphere is written as
\begin{align} 
\phi & = (1 - \alpha) \,\phi(\vec{x},\vec{\zeta} - 2 \zeta_r \vec{n}) 
\nonumber \\
& + \alpha \left[\frac{1+\sigma_{\rmw}}{\pi^{3/2}} 
\exp\left(-\zeta_r^2 - \zeta_\theta^2 - (\zeta_\varphi - \hat{\Omega}_0 \sin 
\theta)^2 \right) E^{-1} - 1 \right],
\quad
\zeta_r > 0, \quad r=1,
\label{e:Maxwell_Bc} 
\end{align}
where the constant $\alpha \in (0,1]$ denotes the accommodation coefficient, $\vec{n}$ (or
$n_i$) the unit normal vector on the sphere pointing to the gas,
$\hat{\Omega}_0 = a \Omega_0/c_0$, and $\sigma_{\rm w}$ is determined by the impermeability condition as follows:
\begin{align}
& 
\sigma_{\rmw} = -2 \sqrt{\pi} \int_{\zeta_r<0} \zeta_r \phi \,E \,\mathrm{d} \bm{\zeta},
\quad r=1.
\label{e:diffuse_sigma}
\end{align} 
Note that the r.h.s of \eqref{e:Maxwell_Bc} is a linear combination of the specular and diffuse reflection conditions. In particular, the Maxwell condition reduces to the diffuse reflection condition when $\alpha=1$.
Finally, we impose the following equilibrium condition at infinity:
\begin{align}
\phi & \to \frac{1}{\pi^{3/2}} \exp\left(-(\zeta_1 - \hat{v}_{\infty 1})^2 
- (\zeta_2 - \hat{v}_{\infty 2})^2 - \zeta_3^2 \right) E^{-1} - 1 
\quad \text{as} \quad r \to \infty,
\label{e:bc_uniform_0}
\end{align}
where $\hat{v}_{\infty i} = v_{\infty i}/c_0$, $i=1,2$.

\subsubsection{Macroscopic quantities} Let $\rho_0(1+ \omega)$, $c_0 u_i$, $T_0(1+\tau)$, 
$p_0 (1+P)$, and $p_0(\delta_{\ij} + P_{\ij})$ $(i,j=1,2,3)$ denote the density, the flow velocity, the temperature, and the stress tensor of the gas, respectively. Here, $\delta_{\ij}$ stands for Kronecker's $\delta$.
The $(\omega,u_i,\tau,P,P_{\ij})$ are given in terms of $\phi$ as follows:
\begin{subequations} \label{e:macroscopic} 
\begin{align}
& \omega = \langle \phi \rangle, 
\quad 
(1 + \omega) u_i = \langle \zeta_i \phi \rangle, 
\quad
\frac{3}{2} (1 + \omega) \tau 
= \left \langle \left(\zeta_j^2 - \frac{3}{2} \right) \phi \right \rangle 
- (1 + \omega) u_j^2, 
 \label{e:macroscopic_a} 
\\
&
P = \omega + \tau + \omega \tau \left(=\frac{P_{11}+P_{22}+P_{33}}{3}\right),
\quad 
P_{\ij} = 2 \langle \zeta_i \zeta_j \phi \rangle 
- 2 (1 + \omega) u_i u_j,
 \label{e:macroscopic_c} 
\end{align} 
\end{subequations}
where the brackets $\langle \,\, \rangle$ stand for
\begin{align} 
\langle g \rangle = \int_{\mathbb{R}^3} g(\bm{\zeta}) E \mathrm{d} \bm{\zeta}.
\end{align}
The stress tensor $P_{\ij}$ is directly related to the force and torque acting on the sphere, as seen from \eqref{e:force_torque_def} below.

Let $F_i$ (or $\bm{F}$) and $M_i$ (or $\bm{M}$) denote the force and torque acting on the sphere, respectively. We introduce their dimensionless counterparts through the relations
\begin{align}
    & \mathcal{F}_i := \frac{F_i}{p_0 a^2},
    \quad
    \mathcal{M}_i := \frac{M_i}{p_0 a^3},
    \quad i=1,2,3.
\end{align}
Note that $\mathcal{F}_i$ (or $\bm{\mathcal{F}}$) and $\mathcal{M}_i$ (or $\bm{\mathcal{M}}$) are expressed in terms of the dimensionless stress by
\begin{align} \label{e:force_torque_def}
& \mathcal{F}_i = - \int_{|\vec{x}|=1} P_{\ij} n_j \,\DS,
\quad
\mathcal{M}_i = - \int_{|\vec{x}|=1} \varepsilon_{\ij m} x_j P_{ml} n_l \,\DS,
\end{align}
where $\DS(=\sin \theta\,\D \theta \D \varphi)$ is the surface element,
$\varepsilon_{\ij k}$ $(i,j,k=1,2,3)$ is Eddington's $\varepsilon$ (Levi--Civita symbol),
and the integration is carried out over the whole surface of the unit sphere $|\vec{x}|=1$.

\subsubsection{Scaling}

The problem \eqref{e:boltzmann}, \eqref{e:Maxwell_Bc}--\eqref{e:bc_uniform_0} is characterized by the following physical parameters:
\begin{align}
    k \ (\text{or} \ \kn),  \quad \hat{\Omega}_0, \quad \hat{\bm{v}}_\infty = (\hat{v}_{\infty 1}, \hat{v}_{\infty 2},0), \quad \alpha.
\end{align}
If we introduce the angle $\alpha_0$ between the vector $\hat{\bm{v}}_\infty$ and  positive $x_1$ axis (see Fig.~\ref{fig:problem}) and write 
$\hat{v}_\infty:= |\hat{\bm{v}}_\infty| = \sqrt{\hat{v}_{\infty 1}^2 + \hat{v}_{\infty 2}^2}$, each component of $\hat{\bm{v}}_\infty$ is expressed as
\begin{align} \label{e:def_parameters_velo}
\hat{v}_{\infty 1} = \hat{v}_\infty \cos \alpha_0, 
\quad \hat{v}_{\infty 2} = \hat{v}_\infty \sin \alpha_0.
\end{align}
Now we recall the assumptions (iii) and (iv) of Sect.~\ref{subsec:problem} and put
\begin{align}
    \epsilon \equiv \hat{v}_\infty \ll 1,
    \quad \hat{\Omega}_0 = S \epsilon,
    \label{e:def_parameters_S}
\end{align}
where the constant $S$ is of the order of unity, i.e.,
$S = \frac{\hat{\Omega}_0}{\epsilon} = \frac{a|\bm{\Omega}_0|}{|\bm{v}_\infty|} 
= O(1)$.
In the asymptotic analysis presented below, we use $S$ instead of $\hat{\Omega}_0$. Further, we assume $k=O(1)$ so that $\epsilon$ plays the role of a sole small parameter.
Note that $k=O(1)$ does not necessarily mean that the value of $k$ is moderate or large; it can be small as long as $k \gg \epsilon$.

In summary, the boundary-value problem to be solved consists of \eqref{e:boltzmann}, \eqref{e:Maxwell_Bc}--\eqref{e:bc_uniform_0} with \eqref{e:def_parameters_velo} and \eqref{e:def_parameters_S}.
In this problem, $\epsilon$ plays the role of a small parameter, and we shall study the asymptotic behavior of the solution when $\epsilon \ll 1$.
In the sequel, we put
\begin{align}
U = \cos \alpha_0, 
\quad V = \sin \alpha_0,
\end{align}
to simplify the notation ($U^2+V^2=1$), and
assign the symbol $\zeta$ to denote $|\bm{\zeta}|=(\zeta_j^2)^{1/2}$.

\section{Asymptotic analysis}
We carry out an asymptotic analysis for small $\epsilon (\ll 1)$ for the problem introduced in the previous section.

We first expand the solution $\phi$ in the form
\begin{align} \label{e:expansion_phi}
    \phi^\epsilon = \phi^{(1)} \epsilon + \phi^{(2)} \epsilon^2 + o(\epsilon^2), \quad \epsilon \ll 1.
\end{align}
The corresponding expansion for each macroscopic quantity is given by
\begin{align} \label{e:expansion_macro}
    h^\epsilon= h^{(1)} \epsilon + h^{(2)} \epsilon^2 + o(\epsilon^2)
    \quad (h = \omega, \ u_i, \ \tau, \, P, \, P_{\ij}).
\end{align}
Note that the relations between $h^{(m)}$ and $\phi^{(m)}$ ($m=1,2$) are obtained by 
substituting \eqref{e:expansion_phi} and \eqref{e:expansion_macro} into \eqref{e:macroscopic} and collecting the terms of the same order.

\subsection{\label{subsec:phi_1}Leading order in $\epsilon$}
The problem for the leading-order term $\phi^{(1)}$ is formally obtained by substituting \eqref{e:expansion_phi} into the original boundary-value problem and retaining the leading-order terms in $\epsilon$, yielding
\begin{align}
& 
\vec{\zeta} \cdot \nabla_x \phi^{(1)}= \frac{1}{k} \mathcal{L}(\phi^{(1)}), 
\quad r>1, 
\label{e:B1} 
\\
& \phi^{(1)} = 
(1-\alpha) \,\phi^{(1)}(\vec{x},\vec{\zeta}-2\zeta_r \vec{n})
+ \alpha \mathcal{K}(\phi^{(1)})
+ I_{\rmw}^{(1)},
\quad \zeta_r > 0, \quad r=1, 
\label{e:dr1} 
\\
& \phi^{(1)} \to I_\infty^{(1)}
\quad \text{as} \quad r \to \infty,
\label{e:inf1} 
\end{align}
with
\begin{align}
    I_{\rmw}^{(1)} & = 2 \alpha S \zeta_\varphi \sin \theta, 
    \quad
    I_\infty^{(1)} = 2(\zeta_1 U + \zeta_2 V),
\end{align}
and
\begin{align}
\mathcal{K}(g) = - 2 \sqrt{\pi} \int_{\zeta_r < 0} \zeta_r g(\bm{\zeta}) E \mathrm{d} \bm{\zeta}.
\end{align}
Equations~\eqref{e:B1}--\eqref{e:inf1} form a boundary-value problem for the linearized Boltzmann equation in an unbounded domain, and its solution is obtained as a superposition of two functions $\phi_\rmU^{(1)}$ and $\phi_\rmS^{(1)}$, i.e.,
\begin{align} \label{e:superposition_1}
\phi^{(1)} = \phi_{\rmU}^{(1)} + \phi_{\rmS}^{(1)},
\end{align} 
where $\phi_\rmU^{(1)}$ and $\phi_\rmS^{(1)}$ are the solutions to \eqref{e:B1}--\eqref{e:inf1} with $I_{\rmw}^{(1)}=0$ and $I_\infty^{(1)}=0$, respectively.
Specifically, $\phi_\rmU^{(1)}$ and $\phi_\rmS^{(1)}$ solve $\vec{\zeta} \cdot \nabla_x \phi^{(1)}_{\rm J} = \frac{1}{k} \mathcal{L}(\phi^{(1)}_{\rm J})$ $({\rm J} = \rmU, \ \rmS)$ under the following respective conditions:
\begin{align}
& \left\{
\begin{array}{ll}
\phi^{(1)}_{\rmU} = (1-\alpha) \,\phi^{(1)}_{\rmU}(\vec{x},\vec{\zeta}-2\zeta_r \vec{n})
+ \alpha \mathcal{K}(\phi^{(1)}_{\rmU}),
\quad \zeta_r > 0, \quad r=1, \\[2mm]
\phi^{(1)}_{\rmU} \to 2(\zeta_1 U + \zeta_2 V) \quad \text{as} \quad r \to \infty,
\end{array}
\right.
\end{align}
and 
\begin{align}
& \left\{
\begin{array}{ll}
\phi^{(1)}_{\rmS} = (1-\alpha) \,\phi^{(1)}_{\rmS}(\vec{x},\vec{\zeta}-2\zeta_r \vec{n})
+ \alpha \mathcal{K}(\phi^{(1)}_{\rmS})
+ 2 \alpha S \zeta_\varphi \sin \theta, 
\quad \zeta_r > 0, \quad r=1, \\[2mm]
\phi^{(1)}_{\rmS} \to 0 \quad \text{as} \quad r \to \infty.
\end{array}
\right.
\end{align}

Note that the problems for $\phi_\rmU^{(1)}$ and $\phi_\rmS^{(1)}$ correspond to classical problems of a rarefied gas flow around a sphere under linearization assumptions.
Indeed, the problem for $\phi_\rmU^{(1)}$ is equivalent to that of a slow flow 
past a stationary sphere \cite{Cercignani-Pagani-Bassanini-PHF-1968,Aoki-Sone-PHF-1987,Sone-Aoki-RGD-1977,Loyalka-PHF-1992,Takata-Sone-Aoki_PHF93,kalempa_sharipov_jfm_2020}\footnote{In the actual analysis, it suffices to consider the particular case $U=1$ and $V=0$ thanks to the similarity solution \eqref{e:similarity_U} introduced below.}. 
In contrast, the problem for $\phi_\rmS^{(1)}$ describes a flow around a rotating sphere in an otherwise stationary gas \cite{Loyalka-PHF-1992,Andreev+Popov_FluidDyn_2010,taguchi_saito_takata_JFM2019}. These problems have been extensively studied in the past few decades (see, e.g., \cite{Sone07,taguchi_saito_takata_JFM2019}).

\subsubsection{\label{subsubsec:similarity}Similarity solutions}

Thanks to the identities summarized in \eqref{e:tensor_fields_L} in Sect.~\ref{sec:analytical_tools} below, we can seek
$\phi^{(1)}_\rmU$ and $\phi^{(1)}_\rmS$ in the forms
\begin{subequations}\label{e:similarity}
\begin{align} 
\phi_{\rmU}^{(1)} & = (U \cos \theta + V \sin \theta \cos \varphi)\, 
\phi_{\rmU a}^{(1)}
\nonumber \\ 
& 
+  
[\zeta_\theta (U \sin \theta - V \cos \theta \cos \varphi) + 
V \zeta_\varphi \sin \varphi]\, \phi_{\rmU b}^{(1)}, 
\label{e:similarity_U}
\\ 
\phi_{\rmS}^{(1)} & = S\, \zeta_\varphi \sin \theta\, \phi_{\rmS a}^{(1)},
\label{e:similarity_S}
\end{align}
\end{subequations}
where $\phi_{\rmU a}^{(1)}$, $\phi_{\rmU b}^{(1)}$, and 
$\phi_{\rmS a}^{(1)}$, which are functions of $(r,\zeta_r,\zeta)$, are unknowns to be determined.
The explicit forms of the problem for $\phi_{\rmU a}^{(1)}$ and $\phi_{\rmU b}^{(1)}$ and that for 
$\phi_{\rmS}^{(1)}$ are easily obtained by direct substitution and will be omitted for conciseness (cf. (3.17) and (3.18) in \cite{taguchi_tsuji_2022_jfm} for the case $\alpha=1$). Note that the number of independent variables is lowered to three. Furthermore, the dependency of $\phi_\rmU^{(1)}$ and $\phi_\rmS^{(1)}$ on the variables $\theta$ and $\varphi$ is explicit in \eqref{e:similarity}.
Consequently, the dependency of each macroscopic quantity on these variables is also explicit. For example, substituting \eqref{e:superposition_1} and \eqref{e:similarity} into \eqref{e:macroscopic}, the components of the leading-order flow velocity are expressed as
\begin{align}\label{e:macro_velo_1}
\left\{\ 
\begin{array}{l}
u_{r}^{(1)} = (U \cos \theta + V \sin \theta \cos \varphi)\, 
\langle \zeta_r \phi_{\rmU a}^{(1)} \rangle,
\\[1.2mm] 
u_{\theta}^{(1)} = (U \sin \theta - V \cos \theta \cos \varphi)\, 
\tfrac{1}{2} \langle (\zeta^2 - \zeta_r^2) \phi_{\rmU b}^{(1)} \rangle,
\\[1.2mm] 
u_{\varphi}^{(1)} = (V \sin \varphi)\, 
\tfrac{1}{2} \langle (\zeta^2 - \zeta_r^2) \phi_{\rmU b}^{(1)} \rangle
+ (S \sin \theta)\, 
\tfrac{1}{2} \langle (\zeta^2 - \zeta_r^2) \phi_{\rmS a}^{(1)} \rangle,
\end{array}
\right.
\end{align}
where each of $\langle \cdots \rangle$ in \eqref{e:macro_velo_1} is a function of $r$.
Similar expressions can be derived for $P_{\ij}^{(1)}$, which are required to compute the force and torque acting on the sphere and thus play a crucial role, as shown next.

\subsubsection{Force and torque on the sphere: order $\epsilon$}
Suppose $\phi_{\rmU a}^{(1)}$, $\phi_{\rmU b}^{(1)}$, $\phi_{\rmS a}^{(1)}$ and thus $P_{\ij}^{(1)}$ are known.
The force and torque acting on the sphere are obtained by substituting the expression of $P_{\ij} = P_{\ij}^{(1)} \epsilon + O(\epsilon^2)$ into \eqref{e:force_torque_def} and carrying out the integration with respect to $\theta$ and $\varphi$,
yielding
\begin{align} \label{e:force_torque_1st}
    \bm{\mathcal{F}} = \hat{\bm{v}}_{\infty} h_D + O(\epsilon^2),
    \quad 
    \bm{\mathcal{M}} = \hat{\bm{\Omega}}_{0} h_M + O(\epsilon^2),
\end{align}
where $\hat{\bm{v}}_\infty = \frac{\bm{v}_\infty}{c_0} = (\hat{v}_{\infty 1}, \hat{v}_{\infty 2},0)$, 
$\hat{\bm{\Omega}}_0 = \frac{a \bm{\Omega}_0}{c_0} = (\hat{\Omega}_0,0,0)$,
and the quantities $h_D$ and $h_M$, independent of $\vec{x}$ and $\vec{\zeta}$, are given in terms of $\phi_{\rmU a}^{(1)}$, $\phi_{\rmU b}^{(1)}$, and $\phi_{\rmS a}^{(1)}$ by
\begin{align} \label{e:definition_hd_hm}
\left\{\
\begin{array}{ll}
\displaystyle
h_D 
= - \lim_{r \to 1} \frac{8}{3} \pi \left( \langle \zeta_r^2 \,\phi_{\rmU a}^{(1)} \rangle -  \langle \zeta_r\, (\zeta^2 - \zeta_r^2)\, \phi_{\rmU b}^{(1)} \rangle \right),
\\[3mm]
\displaystyle
h_M = - \lim_{r \to 1} \frac{8}{3} \pi \langle \zeta_r\, (\zeta^2 - \zeta_r^2) \,\phi_{\rmS a}^{(1)} \rangle.
\end{array}
\right.
\end{align}
Note that $h_D$ (or $h_M$) depends on $k$ and $\alpha$ through $\phi_{\rmU a}^{(1)}$ and $\phi_{\rmU b}^{(1)}$ (or $\phi_{\rmS a}^{(1)}$).
Numerical values of $h_D$ for various $k$ can be found in \cite{Takata-Sone-Aoki_PHF93,Sone07,Taguchi+Suzuki_PhysRevFluids_2017} in the case of $\alpha=1$, while those of $h_M$ are tabulated in \cite{taguchi_saito_takata_JFM2019} for various $k$ and $\alpha$.

According to \eqref{e:force_torque_1st}, the leading-order force acting on the sphere is parallel to $\vec{v}_\infty$, indicating the transverse force is absent at this stage. The absence of transverse force in \eqref{e:force_torque_1st} is due to the lack of coupling between the uniform and rotating flows in the linearized system. (No coupling terms between $\phi_\rmU^{(1)}$ and $\phi_\rmS^{(1)}$ occur in the system.) 
In the following subsection, therefore, we proceed to the second-order problem for $\phi^{(2)}$ to study the coupling effect. 

We conclude this subsection with a brief comment on the physical interpretation of $h_D$ and $h_M$.
Writing the force and torque obtained above using the original dimensional quantities, we have
\begin{align} \label{e:stokes_kirchhoff}
    \bm{F} &= 6 \pi \mu a \bm{v}_\infty \bar{h}_D,
    \quad
    \bm{M} = -8 \pi \mu a^3 \bm{\Omega}_0 \bar{h}_M,
\end{align}
where $\mu = \frac{\sqrt{\pi}}{2} \frac{\gamma_1 p_0 \ell_0}{c_0}$ is the viscosity at the reference state \cite{Sone07} and
\begin{align}
    \bar{h}_D = \frac{h_D}{6 \pi \gamma_1 k},
    \quad
    \bar{h}_M = - \frac{h_M}{8\pi \gamma_1 k}.
\end{align}
Here, the dimensionless viscosity constant
$\gamma_1 >0$ is 
given by $\gamma_1 = \frac{2}{15} \langle \zeta^4 B \rangle$, where
$B=B(\zeta)$ is the solution to the integral equation
\begin{align}
\mathcal{L} (\zeta_{\ij} B )=-2 \zeta_{\ij},
\quad \zeta_{\ij} = \zeta_i \zeta_j - \tfrac{1}{3}\zeta^2 \delta_{\ij}.
\end{align}
The value of $\gamma_1$ for the BGK model is the unity ($\gamma_1=1$), while that for a hard-sphere gas is $\gamma_1 = 1.270042427$ \cite{Sone07}.
Equation~\eqref{e:stokes_kirchhoff} coincides with the Stokes and Kirchhoff formulas for the drag and torque acting on a sphere if $\bar{h}_D = \bar{h}_M=1$.
Therefore, we can interpret $\bar{h}_D$ (or $h_D$) and $\bar{h}_M$ (or $h_M$) as numerical factors that represent the deviations from the Stokes and Kirchhoff formulas, respectively.
Note that $\bar{h}_D \to 1$ and $\bar{h}_M \to 1$ as $k \to 0$ \footnote{The asymptotic expressions for $\bar{h}_D$ and $\bar{h}_M$ for $k\ll1$ are obtained as $\bar{h}_D = 1+k_0\,k + O(k^2)$ \cite{Sone07} and $\bar{h}_M = 1+3k_0\,k + O(k^2)$ \cite{taguchi_saito_takata_JFM2019}. See the caption of Fig.~\ref{fig:fig_hl_vs_kn_bgk} for the meaning of $k_0$.}. Therefore,
we recover the Stokes and Kirchhoff formulas for the drag and torque as $k \to 0$ ($\epsilon \ll k$).

\subsection{Second order in $\epsilon$}

We consider the second-order problem.

\subsubsection{Asymptotic matching}
In Sect.~\ref{subsec:phi_1}, we have discussed the leading-order term $\phi^{(1)}( = \phi_{\rmU}^{(1)} + \phi_{\rmS}^{(1)})$ based on the linearized Boltzmann equation.
In \cite{Taguchi-JFM2015}, it has been pointed out that $\phi_{\rmU}^{(1)}$ does not approximate the solution of the Boltzmann equation uniformly in space. For clarity, let us consider the case $U=1$ and $V=0$. 
The approach of $\phi_{\rmU}^{(1)}$ to $I_\infty^{(1)} = 2 \zeta_1$ as $r \to \infty$ is proportional to
$r^{-1}$ \cite{Takata-Sone-Aoki_PHF93,Taguchi-JFM2015}.
Using this information,
the magnitude of the streaming term of the Boltzmann equation is estimated as $\vec{\zeta} \cdot \nabla_x (\epsilon \phi_{\rmU}^{(1)}) \sim \epsilon/r^2$ for $r = |\vec{x}| \gg 1$. On the other hand, we can transform the nonlinear term as $\mathcal{J}(\epsilon \phi_{\rmU}^{(1)},\epsilon \phi_{\rmU}^{(1)}) \sim - \epsilon^2 \mathcal{L}(2\zeta_1^2) + O(\epsilon^2/r)$, where the term $O(\epsilon^2/r)$ contains nonlinear terms. 
Thus, the nonlinear term decays slower than the streaming term, eventually becoming comparable to it for large $r$, no matter how small $\epsilon$ is. 
In other words, the term $\varepsilon \phi^{(1)}$ based on the linearized equation does not serve as a good approximate solution to the original system in the region far from the sphere (owing to the term $\phi_{\rmU}^{(1)}$)%
\footnote{We encounter similar shortcomings of a linearized system when considering low Reynolds number flows past a sphere based on the Navier--Stokes equation (Whitehead's paradox \cite{Van-Dyke-1975}).}.
To overcome this difficulty, we amend the simple expansion \eqref{e:expansion_phi} to consider a different length scale in the far field. Hence, we introduce the notion of a slowly varying solution and start over the analysis using the method of matched asymptotic expansion developed in \cite{Taguchi-JFM2015}.

Following \cite{Taguchi-JFM2015,taguchi_tsuji_2022_jfm},
we consider two regions, the inner ($1 < r \ll \epsilon^{-1}$) and outer ($1 \ll r < \infty$) regions, overlapping in the intermediate region ($1 \ll r \ll \epsilon^{-1}$). 
Hereafter, we denote by $\phi_{\rmH}$ the solution in the outer region and regard the solution in \eqref{e:expansion_phi} as a solution in the inner region, valid in the region $1 < r \ll \epsilon^{-1}$. We assume that the length scale of variation of the solution in the outer region is of the order of $1/\epsilon$ (\textit{slowly varying solution}). In other words, $\partial \phi_{\rmH}/\partial x_i = O(\epsilon \phi_{\rmH})$ (except possibly for the intermediate region).
The following analysis is consistent with these assumptions.

We assume that the slowly varying solution can be expanded as
\begin{align} \label{e:expansion_outer}
    \phi_{\rmH}^\epsilon = \epsilon \phi_{\rmH}^{(1)} + \epsilon^2 \phi_{\rmH}^{(2)} + o(\epsilon^2).
\end{align}
Since $\phi_{\rmH}$'s length scale of variation is of the order of $\epsilon^{-1}$, we can identify this expansion with a Hilbert expansion for finite Reynolds numbers (S expansion \cite{Sone07}).
Consequently, the solution in the outer region is described \textit{fluid-dynamically} by a set of incompressible Navier-Stokes-type equations, whose explicit forms are shown in (3.36)--(3.38) of \cite{taguchi_tsuji_2022_jfm}.
Using this fact, as well as the asymptotic behavior of $\phi^{(1)}$ in the far field,
we can determine the first two terms 
of the expansion \eqref{e:expansion_outer}\footnote{To be specific, $\phi_{\rmH}^{(1)} = 2(U \zeta_1 + V \zeta_2)$ and $\phi_{\rmH}^{(2)}$ is given by a Maxwellian with corresponding macroscopic flow velocity given by an Oseenlet (see (3.59) in \cite{taguchi_tsuji_2022_jfm}).
}.
Once $\phi_{\rmH}^\epsilon$ is determined to order $\epsilon^2$, we can derive a matching condition for $\phi^{(2)}$ at $r \to \infty$, which serves as a boundary condition at infinity. Since the process of derivation is the same as that in \cite{taguchi_tsuji_2022_jfm}, we omit further details and concentrate on the derived problem for $\phi^{(2)}$ next.

\subsubsection{Problem for $\phi^{(2)}$}

$\phi^{(2)}$ is determined through the following problem:
\begin{align}
& 
\vec{\zeta} \cdot \nabla_x \phi^{(2)}= \frac{1}{k} \mathcal{L}(\phi^{(2)}) + \frac{1}{k} I^{(2)}, 
\quad r>1,
\label{e:LB_inner_order2} \\
& \phi^{(2)} =
(1-\alpha) \,\phi^{(2)}(\vec{x},\vec{\zeta}-2\zeta_r \vec{n})
+ \alpha \mathcal{K}(\phi^{(2)})
+ I_{\rmw}^{(2)},
\quad \zeta_r > 0, \quad r = 1,
\label{e:inner_order2_BC}
\\
& \phi^{(2)} \to I_\infty^{(2)}
\quad
\text{as} \quad r \to \infty,
 \label{e:inner_order2_MC}
\end{align}
with
\begin{subequations}
\begin{align}
    I^{(2)} & = \mathcal{J}(\phi^{(1)},\phi^{(1)}),
    \label{e:inhomogeneous_2}
    \\
    I_{\rmw}^{(2)} & = \alpha \,(2 \zeta_\varphi^2 - 1) \,S^2 \sin^2 \theta 
+ 2 \alpha \mathcal{K}(\phi^{(1)}) \,\zeta_\varphi \,S \sin \theta,
\label{e:inhomogeneous_2_Iw}
\\
    I_\infty^{(2)} & = 2 (U \zeta_1 + V \zeta_2)^2 - 1
    \nonumber \\
    &
    + \frac{3\, \bar{h}_D}{2 \gamma_1 k}
    \bigg(- \frac{1}{2} \zeta_r (U \cos \theta + V \sin \theta \cos \varphi
- 1) 
\,
(3(U \cos \theta + V \sin \theta \cos \varphi) + 1)
\nonumber 
\\
& 
\qquad + \zeta_\theta (U \sin \theta - V \cos \theta \cos
\varphi)
\,(U \cos \theta + V \sin \theta \cos \varphi - 1)
\nonumber \\
& 
\qquad + \zeta_\varphi V \sin \varphi
\,(U \cos \theta + V \sin \theta \cos \varphi - 1) \bigg).
\label{e:inhomogeneous_inf}
\end{align}
\end{subequations}
The condition \eqref{e:inner_order2_MC} with \eqref{e:inhomogeneous_inf} is required for the inner solution $\phi$ to match the outer solution $\phi_{\rmH}$.
The physical meaning of the term proportional to $\bar{h}_D$ in $I_\infty^{(2)}$ is a correction due to the long-range integration of the nonlinear term, omitted in the analysis of $\phi_{\rmU}^{(1)}$. The first two terms of \eqref{e:inhomogeneous_inf} is a contribution from the uniform flow distribution \eqref{e:bc_uniform_0} at infinity.

\subsubsection{\label{subsubsec:similarity_2}Similarity solutions}

The problem \eqref{e:LB_inner_order2}--\eqref{e:inner_order2_MC} is similar to the problem for $\phi^{(1)}$ except for the inhomogeneous terms, which suggests looking for a similarity solution for this problem.

We first put
\begin{align} \label{e:inner_second_order}
\phi^{(2)} 
& = 
\frac{3\,\bar{h}_D}{4 \gamma_1 k} \,(2 U \zeta_1 + 2 V \zeta_2)
\,\phi_{\rmU}^{(1)} + \phi_{\rmUU}^{(2)} + \phi_{\rmSS}^{(2)} + \phi_{\rmUS}^{(2)},
\end{align}
where $\phi_{\rmUU}^{(2)}$, $\phi_{\rmSS}^{(2)}$, and $\phi_{\rmUS}^{(2)}$ are functions to be determined. The
$\phi_{\J}^{(2)}$ ($\J=\rmUU,\rmSS,\rmUS$) satisfies
\eqref{e:LB_inner_order2}--\eqref{e:inner_order2_MC} with
$\phi^{(2)} = \phi_{\J}^{(2)}$, $I^{(2)} = I_{\J}^{(2)}$, $I_{\rmw}^{(2)} = I_{\rmw,\J}^{(2)}$, $I_\infty^{(2)} = I_{\infty,\J}^{(2)}$, where
\begin{align}
& I_{\rmUU}^{(2)} = \mathcal{J}(\phi_{\rmU}^{(1)},\phi_{\rmU}^{(1)}), 
\quad
I_{\rmSS}^{(2)} = \mathcal{J}(\phi_{\rmS}^{(1)},\phi_{\rmS}^{(1)}),
\quad
I_{\rmUS}^{(2)} = 2\mathcal{J}(\phi_{\rmU}^{(1)},\phi_{\rmS}^{(1)}), 
\label{e:iht_order_2}
\\[2mm]
&
I_{{\rm w},\rmUU}^{(2)} = 0,
\quad 
I_{{\rm w},\rmSS}^{(2)} = \alpha (2 \zeta_\varphi^2 - 1) S^2 \sin^2 \theta,
\quad
I_{{\rm w},\rmUS}^{(2)} = 
2 \alpha \mathcal{K}(\phi_{\rmU}^{(1)}|_{r=1})
\,\zeta_\varphi \,S \sin \theta,
\label{e:iht_w_order_2}
\\
& I_{\infty,\rmUU}^{(2)} = (U \cos \theta + V \sin \theta \cos \varphi)^2 
\left(
-\frac{9}{4} \frac{\bar{h}_D}{\gamma_1 k} 
\zeta_r + 3 \,\zeta_r^2 - \zeta^2
\right)
\nonumber \\
& \qquad \quad + 
(U \cos \theta + V \sin \theta
\cos \varphi) 
\left[(U \sin \theta - V \cos \theta \cos \varphi) \,\zeta_\theta + V \sin \varphi
\,\zeta_\varphi \right]
\nonumber \\
& \qquad \quad \times
\left(
\frac{3\,\bar{h}_D}{2 \gamma_1 k} 
- 4 \zeta_r \right)
\nonumber \\
& \qquad \quad + 2 \bigg\{
\left[(U \sin \theta - V \cos \theta \cos \varphi)^2
- V^2 \sin^2 \varphi \right] \frac{\zeta_\theta^2 - \zeta_\varphi^2}{2}
\nonumber \\
& \qquad \quad
+ 2 V \sin \varphi \,(U \sin \theta - V \cos \theta \cos \varphi) \,\zeta_\theta
\,\zeta_\varphi \bigg\}
+ \frac{3\,\bar{h}_D}{4 \gamma_1 k} 
\zeta_r + \zeta^2 - \zeta_r^2 - 1,
\\
& I_{\infty,\rmSS}^{(2)} = I_{\infty,\rmUS}^{(2)} = 0.
\label{e:I_infUU}
\end{align}

We want to find a similarity solution for each of $\phi_{\J}^{(2)}$.
For this purpose, we note that not only $\mathcal{L}$ but also $\mathcal{J}$ is an isotropic operator and use the identities shown in Sect.~\ref{subsec:isotropy} below. 
Here,
we only show the form of a similarity solution for $\phi_{\rmUS}^{(2)}$ \cite{taguchi_tsuji_2022_jfm} because only $\phi_{\rmUS}^{(2)}$ contributes to the transverse force, among others: 
\begin{align}
\phi_{\rmUS}^{(2)} & = \phi_{\rmUS}^{(2)\sharp} + \phi_{\rmUS}^{(2)\flat},
\label{e:similarity_sol_US}
\\
\phi_{\rmUS}^{(2)\sharp} & =
S \sin \theta \,(U \cos \theta + V \sin \theta \cos \varphi)
\,\zeta_\varphi \,\phi_{\rmUS a}^{(2)\sharp}
\nonumber \\
& + S \cos \theta \left[ (U \sin \theta - V \cos \theta \cos \varphi) \,\zeta_\varphi -
V \sin \varphi \,\zeta_\theta \right] \phi_{\rmUS b}^{(2)\sharp}
\nonumber \\
& + S \sin \theta \left[ -V \sin \varphi \frac{\zeta_\theta^2 - \zeta_\varphi^2}{2} 
 +
 (U \sin \theta - V \cos \theta \cos \varphi) \,\zeta_\theta \,\zeta_\varphi \right]
 \phi_{\rmUS c}^{(2)\sharp}
\nonumber \\
& + SV \sin \theta \sin \varphi \,\phi_{\rmUS d}^{(2)\sharp},
\label{e:similarity_sol_US_1}
\\
\phi_{\rmUS}^{(2)\flat} & =
S \sin \theta \,(U \sin \theta - V \cos \theta \cos \varphi)
\,\phi_{\rmUS a}^{(2)\flat}
\nonumber \\
& + S \left[ (U \sin 2 \theta - V \cos 2\theta \cos \varphi) \,\zeta_\theta +
 V \cos \theta \sin \varphi \,\zeta_\varphi \right] \phi_{\rmUS b}^{(2)\flat}
\nonumber \\
& + S \sin \theta \left[ (U \sin \theta - V \cos \theta \cos \varphi) \frac{\zeta_\theta^2 - \zeta_\varphi^2}{2} 
+ V \sin \varphi \,\zeta_\theta \,\zeta_\varphi \right]
\phi_{\rmUS c}^{(2)\flat}
\nonumber \\
& + SU \phi_{\rmUS d}^{(2)\flat},
\label{e:similarity_sol_US_2}
\end{align}
where $\phi_{\rmUS \beta}^{(2)\sharp}$ and $\phi_{\rmUS \beta}^{(2)\flat}$ ($\beta=a,b,c,d$) on the r.h.s. of \eqref{e:similarity_sol_US_1} and \eqref{e:similarity_sol_US_2} are functions of $r$, $\zeta_r$, and $\zeta$, whose dependency is not shown explicitly.
Note that the similarity solutions for $\phi_{\J}^{(2)}$ are the same as those for $\alpha=1$ used in \cite{taguchi_tsuji_2022_jfm} (cf.~(3.79)--(3.83) there).

The equations and boundary conditions for $\phi_{\rmUS \beta}^{(2)\sharp}$ and $\phi_{\rmUS \beta}^{(2)\flat}$ are derived once we substitute \eqref{e:similarity_sol_US}--\eqref{e:similarity_sol_US_2} into those for $\phi_{\rmUS}^{(2)}$ and use the identities \eqref{e:tensor_fields_L} and \eqref{e:tensor_fields_J} in Sect.~\ref{subsec:isotropy}.
Since the derived equations and boundary conditions are lengthy, we do not present the explicit forms here (see Appendix~E of \cite{taguchi_tsuji_2022_jfm} for the case $\alpha=1)$.
Note that thanks again to the similarity solutions, we can explicitly write the dependency of the second-order macroscopic quantities on the variables $\theta$ and $\varphi$, as in \eqref{e:macro_velo_1}.

\subsubsection{Force and torque on the sphere: order $\epsilon^2$}

We now discuss general expressions for the force and torque to the second-order approximation in $\epsilon$. The derivation is similar to that for the first order. Therefore, we only summarize the result. First, the second-order term of the torque turns out to be zero, allowing us to write the second equation of \eqref{e:force_torque_1st} as
\begin{align} \label{e:formula_torque}
    \bm{\mathcal{M}} & = \hat{\bm{\Omega}}_{0} h_M + o(\epsilon^2).
\end{align}
Next, 
the force acting on the sphere (the first equation of \eqref{e:force_torque_1st}) improves to
\begin{align}
\bm{\mathcal{F}} & =  \underbrace{\hat{\bm{v}}_{\infty} \left(1 + \epsilon  \frac{3}{4} \frac{\bar{h}_D}{\gamma_1 k} \right) h_D}_{\text{drag}}
+ \underbrace{(\hat{\bm{v}}_{\infty} \times \hat{\bm{\Omega}}_0) \,h_L}_{\text{transverse force}}
+ o(\epsilon^2),
\label{e:formula_force}
\end{align}
where $h_L$ is the quantity defined by 
\begin{align}
& h_L = \lim_{r \to 1} \frac{4}{3} \pi \left(\langle \zeta_r (\zeta^2-\zeta_r^2 ) (\phi_{\rmUS a}^{(2)\sharp} - \phi_{\rmUS b}^{(2)\sharp}) \rangle
+ 2 \langle \zeta_r^2 \phi_{\rmUS d}^{(2)\sharp} \rangle \right).
\label{e:definition_hL}
\end{align}
Note that $h_L$ depends on $k$ and $\alpha$ through $\phi_{\rmUS a}^{(2)\sharp}$, $\phi_{\rmUS b}^{(2)\sharp}$, and $\phi_{\rmUS d}^{(2)\sharp}$. 
In this way, $h_L$ is a quantity representing the effects of gas rarefaction and surface accommodation on the transverse force.

Finally, it is worth expressing the drag, lift, and torque in dimensional forms. Omitting the terms of $o(\epsilon^2)$, they are summarized as
\begin{align}
    & \text{Drag:} && \bm{F}_D = 6 \pi \mu a \bm{v}_\infty \left(1 + \frac{3}{8} \re\, \bar{h}_D \right)\bar{h}_D,
    \label{e:drag_dimensioal}
    \\
    & \text{Lift:} && \bm{F}_ L = \pi \rho_0 a^3 (\bm{v}_\infty \times \bm{\Omega}_0) \,\bar{h}_L,
    \label{e:lift_dimensional}
    \\
    & \text{Torque:} && \bm{M} = -8 \pi \mu a^3 \bm{\Omega}_0 \,\bar{h}_M.
\end{align}
where we have written
\begin{align}
    \bar{h}_L = \frac{h_L}{2\pi},
\end{align}
and $\re = \frac{\rho_0 |\vec{v}_\infty| a}{\mu}$ is the Reynolds number of the flow.
Note the similarities of \eqref{e:drag_dimensioal} and \eqref{e:lift_dimensional} to the Oseen drag for a sphere and to the lift force for a rotating sphere derived by Rubinow \& Keller \cite{Rubinow-Keller_JFM61} for continuum flows, respectively.


\section{Numerical results for $h_L$ and discussions}

In this section, we compute the numerical values of $h_L$ for various $k$ and $\alpha$ based on the BGK model of the Boltzmann equation.

\subsection{Comments on the computation of $h_L$}

The computation of $h_L$ requires the information on $\phi_{\rmUS}^{(2)\sharp}$ (or $\phi_{\rmUS \beta}^{(2)\sharp}$) according to \eqref{e:definition_hL}. Due to the source term that depends on the lower-order solutions, the computation of $\phi_{\rmUS \beta}^{(2)\sharp}$ is numerically quite challenging.
Fortunately, there is a way to bypass this difficulty, devised in \cite{taguchi_tsuji_2022_jfm}, and we shall take the same approach in this study. 

More specifically, in \cite{taguchi_tsuji_2022_jfm}, an alternative formula for $h_L$ has been derived in the case of the diffuse reflection condition ($\alpha=1$). In this study, we have obtained the corresponding formula in the present case of the Maxwell condition. The following formula gives the result for the Boltzmann equation:
\begin{align} \label{e:cross_coupling_formula_hL}
h_L & = \frac{4}{3} \pi \left[
\mathcal{K}(\phi_{\rmU a}^{(1)}|_{r=1}) \,
\langle \zeta_r (\zeta^2-\zeta_r^2) \phi_{\rmU b}^{(1)}|_{r=1} \rangle
\right. \nonumber
\\
&
\left.
 - \frac{1}{k} \sum_{t=\theta,\varphi} \int_1^\infty r^2 \left(\langle \phi_{\rmU a}^{(1)-} \,\mathcal{J}(\zeta_t \phi_{\rmU b}^{(1)},\zeta_t \phi_{\rmS a}^{(1)})
 \rangle
 - \langle \zeta_t \phi_{\rmU b}^{(1)-} \,\mathcal{J}(\phi_{\rmU a}^{(1)}, \zeta_t \phi_{\rmS a}^{(1)} ) \rangle \right) \mathrm{d} r\right]
\end{align}
($0 < k < \infty$, $0 < \alpha \le 1$). Here,
the symbol $^-$ stands for the inversion of the velocity variables, i.e., $\psi^{-}(r,\zeta_r,\zeta) = \psi(r,-\zeta_r,\zeta)$ for
$\psi = \phi_{\rmU \beta}^{(1)}$, $\beta=a,b$, etc.
Note that in \eqref{e:cross_coupling_formula_hL}, $h_L$ is expressed in terms of the lower-order functions $\phi_{\rmU a}^{(1)}$, $\phi_{\rmU b}^{(1)}$, and $\phi_{\rmS a}^{(1)}$, enabling us to compute $h_L$ without solving any second-order problems. The derivation of \eqref{e:cross_coupling_formula_hL} uses the identity derived in \cite{Takata09-1} (see Sect.~\ref{sec:analytical_tools} below).
Interestingly, the above formula does not explicitly depend on $\alpha$ but does so through $\phi_{\rmU a}^{(1)}$, $\phi_{\rmU b}^{(1)}$, and $\phi_{\rmS a}^{(1)}$. Indeed, the formula \eqref{e:cross_coupling_formula_hL}, applicable for any $\alpha \in (0,1]$, is identical with that for $\alpha=1$ derived in \cite{taguchi_tsuji_2022_jfm}. For the BGK model, we can derive the corresponding formula easily from \eqref{e:cross_coupling_formula_hL} by taking into account the corresponding expressions of $\mathcal{J}(\cdot,\cdot)$ for the BGK model (see (4.3), (4.8), and (4.12) of \cite{taguchi_tsuji_2022_jfm}).

We carry out actual computations using the BGK model as in \cite{taguchi_tsuji_2022_jfm}.
To apply the formula \eqref{e:cross_coupling_formula_hL}, we require numerical data for $(\phi_{\rmU a}^{(1)},\phi_{\rmU b}^{(1)})$ and $\phi_{\rmS a}^{(1)}$, which are the solutions to the problem of a uniform flow past a sphere (say, problem U) and that of a swirling flow around a sphere (say, problem S), respectively.
Numerical analysis for problem S under the Maxwell boundary condition has been carried out in \cite{taguchi_saito_takata_JFM2019} for various $\alpha$ and $k$. In the present study, we also made additional computations to obtain data for different $\alpha$. Concerning problem U under the Maxwell boundary condition, we performed calculations using the numerical code in \cite{Taguchi+Suzuki_PhysRevFluids_2017}, where the diffuse reflection condition is assumed, with slight and necessary modifications. 
The reader is referred to \cite{taguchi_tsuji_2022_jfm,Takata-Sone-Aoki_PHF93,Taguchi+Suzuki_PhysRevFluids_2017} for further details.

\subsection{Transverse force acting on a rotating sphere}
\begin{figure}[t]
    \centering
    \includegraphics[width=0.7\linewidth]{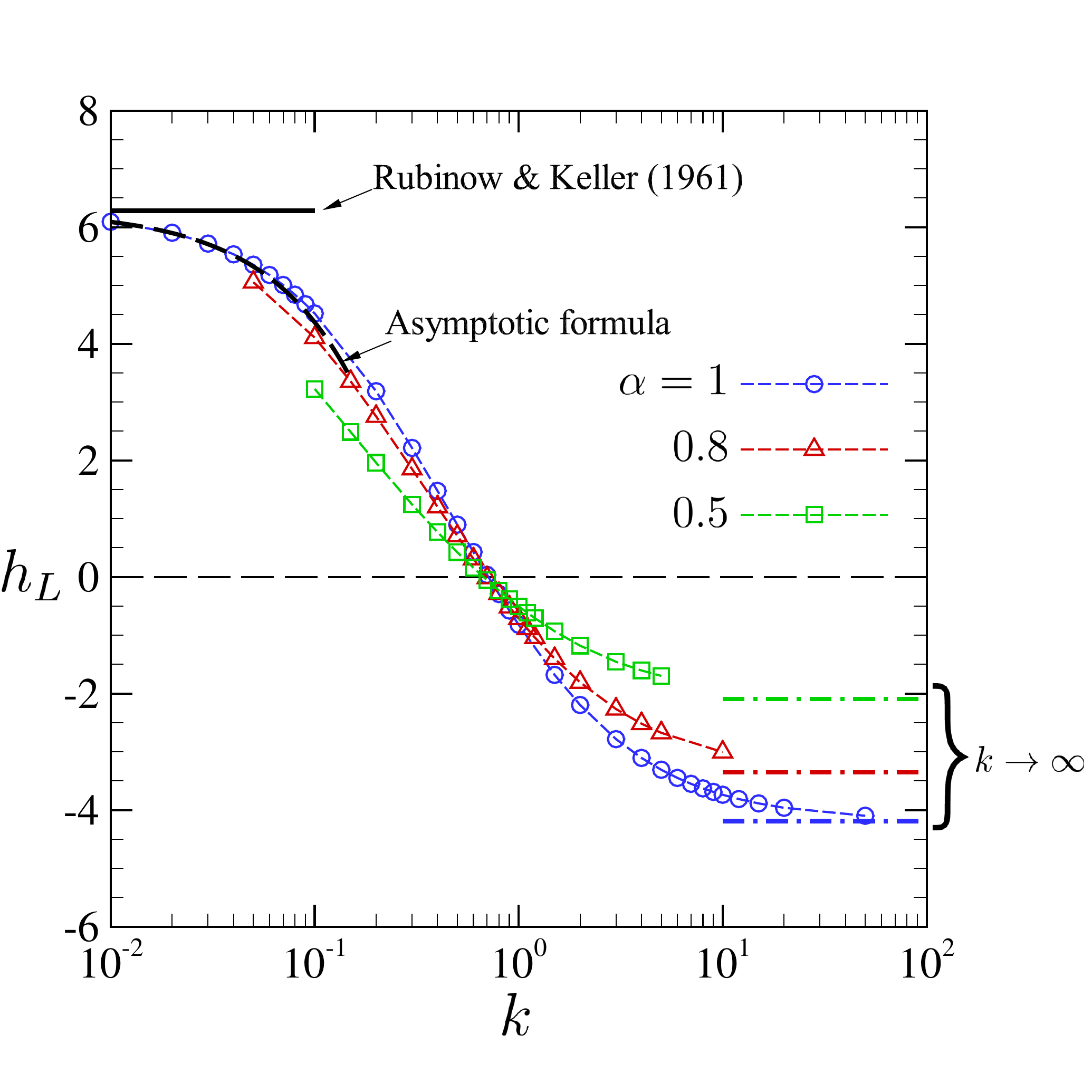}
    \caption{$h_L$ vs $k$ for the BGK model under the Maxwell boundary condition for $\alpha=0.8$ (red triangle) and 0.5 (green square). The results for the case $\alpha=1$ (the diffuse reflection) \cite{taguchi_tsuji_2022_jfm} are shown by the blue circles. The dash-dotted lines indicate the values in the free molecular limit. The dashed curve shows the asymptotic formula $h_L = 2\pi (1 + 3k_0\,k)$ for $k \ll 1$ ($\alpha = 1$) \cite{taguchi_tsuji_2022_jfm}, where $k_0$ is the slip coefficient, whose numerical value is $k_0=-1.01619$ for the BGK model under the diffuse reflection condition \cite{Sone07}. The solid horizontal line indicates the limiting value as $k \to 0$ ($0 < \alpha \le 1$), corresponding to the value for a continuum flow \cite{Rubinow-Keller_JFM61}.}
    \label{fig:fig_hl_vs_kn_bgk}
\end{figure}
Figure~\ref{fig:fig_hl_vs_kn_bgk} shows the obtained $h_L$ as a function of $k$
for $\alpha=1$, 0.8, and 0.5. The numerical data for $\alpha=0.8$ and 0.5 are new and those for $\alpha=1$ are taken from \cite{taguchi_tsuji_2022_jfm}. The corresponding numerical values are tabulated in Table~\ref{table:h_L}, where the values of $h_D$ for $\alpha=0.8$ and 0.5 and those of $h_M$ for $\alpha=0.5$ are also included.
As seen from the figure, the overall tendency of $h_L$ in terms of $k$ is the same for all values of $\alpha$. That is, $h_L$ is decreasing with $k$. For each $\alpha$, there is a threshold in $k$, above which $h_L$ becomes negative. Consequently, the sphere begins to experience a negative lift when $k$ exceeds this threshold (\textit{inverse Magnus effect}). The values of $h_L$ are significantly affected by $\alpha$, especially when $k$ is large. The transverse force tends to become small for smaller $\alpha$. On the other hand, interestingly, the value of $k$ at which $h_L$ becomes zero does not depend much on $\alpha$.
For example, the threshold is 0.710 \cite{taguchi_tsuji_2022_jfm}, 0.695, and 0.674 for $\alpha=1$, 0.8, and 0.5, respectively, for the present computations based on the BGK model.
Incidentally, the limiting value of $h_L$ as $k\to \infty$, i.e., the free molecular limit, is known to be given by $h_L = - \frac{4}{3} \alpha \pi$ \cite{Wang_AIAA_1972,Ivanov-Yanshin_FD80,Borg-Soederholm-Essen-PHF03}.

\def\colsp{0.8em}
\def\sm{\phantom{-}}
\def\sa{\phantom{(0.0000)^a}}
\def\sp{\phantom{0}}
\def\sd{\phantom{.00}}
\def\rp{0.5em}
\def\rowsm{-0.5em}

\begin{table}[tb]\centering
\caption{Values of $h_L$ for various $k$ for the BGK model under the Maxwell boundary condition with accommodation coefficient $\alpha$. 
The values for the case $\alpha=1$ are taken from \cite{taguchi_tsuji_2022_jfm}, in which other values for different $k$ are also available. Values of $h_D$ for $\alpha=0.8$, $0.5$ and those of $h_M$ for $\alpha=0.5$ are also shown, since they are newly reported here. 
}
{\tabcolsep = \colsp
\begin{tabular}{cccccccccc}
  \hline\hline\\[\rowsm]
 & \multicolumn{3}{c}{$h_L$}
 &&\multicolumn{2}{c}{$h_D$}
 &&\multicolumn{1}{c}{$h_M$}\\[\rp]
 \cline{2-4}\cline{6-7}\cline{9-9}\\
 $k        $ & $\alpha=1  $ &  $\alpha=0.8$ &  $\alpha=0.5$ & \, & $\alpha=0.8$ &  $\alpha=0.5$ & \, & $\alpha=0.5$\\[\rp]
 $\sp0.05  $ & $\sm 5.3563$ &  $\sm 5.0578$ &          ---  & \, & $\sp 0.8782$ &          ---  & \, &      --- \\
 $\sp0.1\sp$ & $\sm 4.5204$ &  $\sm 4.1055$ &  $\sm 3.2254$ & \, & $\sp 1.6579$ &  $\sp 1.5554$ & \, & $-1.3218$\\
 $\sp0.15  $ &      ---     &  $\sm 3.3558$ &  $\sm 2.4873$ & \, & $\sp 2.3635$ &  $\sp 2.2142$ & \, & $-1.5830$\\
 $\sp0.2\sp$ & $\sm 3.1866$ &  $\sm 2.7545$ &  $\sm 1.9594$ & \, & $\sp 2.9957$ &  $\sp 2.7947$ & \, & $-1.7489$\\
 $\sp0.3\sp$ & $\sm 2.2129$ &  $\sm 1.8523$ &  $\sm 1.2431$ & \, & $\sp 4.0824$ &  $\sp 3.7905$ & \, & $-1.9420$\\
 $\sp0.4\sp$ & $\sm 1.4763$ &  $\sm 1.2018$ &  $\sm 0.7682$ & \, & $\sp 4.9737$ &  $\sp 4.6052$ & \, & $-2.0478$\\
 $\sp0.5\sp$ & $\sm 0.8970$ &  $\sm 0.7048$ &  $\sm 0.4229$ & \, & $\sp 5.7114$ &  $\sp 5.2778$ & \, & $-2.1132$\\
 $\sp0.6\sp$ & $\sm 0.4272$ &  $\sm 0.3096$ &  $\sm 0.1569$ & \, & $\sp 6.3281$ &  $\sp 5.8388$ & \, & $-2.1571$\\
 $\sp0.7\sp$ & $\sm 0.0362$ &  $   -0.0147$ &  $   -0.0564$ & \, & $\sp 6.8500$ &  $\sp 6.3124$ & \, & $-2.1885$\\
 $\sp0.8\sp$ & $   -0.2935$ &  $   -0.2854$ &  $   -0.2316$ & \, & $\sp 7.2948$ &  $\sp 6.7154$ & \, & $-2.2118$\\
 $\sp0.9\sp$ & $   -0.5762$ &  $   -0.5156$ &  $   -0.3788$ & \, & $\sp 7.6780$ &  $\sp 7.0620$ & \, & $-2.2299$\\
 $\sp1\sd$   & $   -0.8213$ &  $   -0.7140$ &  $   -0.5044$ & \, & $\sp 8.0110$ &  $\sp 7.3628$ & \, & $-2.2442$\\
 $\sp1.1\sp$ &      ---     &  $   -0.8868$ &  $   -0.6129$ & \, & $\sp 8.3027$ &  $\sp 7.6258$ & \, & $-2.2558$\\
 $\sp1.2\sp$ &      ---     &  $   -1.0387$ &  $   -0.7076$ & \, & $\sp 8.5599$ &  $\sp 7.8575$ & \, & $-2.2654$\\
 $\sp1.5\sp$ & $   -1.6807$ &  $   -1.4012$ &  $   -0.9316$ & \, & $\sp 9.1754$ &  $\sp 8.4111$ & \, & $-2.2862$\\
 $\sp2\sd$   & $   -2.1964$ &  $   -1.8082$ &  $   -1.1797$ & \, & $\sp 9.8661$ &  $\sp 9.0305$ & \, & $-2.3066$\\
 $\sp3\sd$   & $   -2.7827$ &  $   -2.2667$ &  $   -1.4554$ & \, & $   10.6394$ &  $\sp 9.7220$ & \, & $-2.3263$\\
 $\sp4\sd$   & $   -3.1052$ &  $   -2.5172$ &  $   -1.6046$ & \, & $   11.0581$ &  $   10.0955$ & \, & $-2.3359$\\
 $\sp5\sd$   & $   -3.3085$ &  $   -2.6746$ &  $   -1.6978$ & \, & $   11.3194$ &  $   10.3283$ & \, & $-2.3415$\\
 $  10\sd$   & $   -3.7368$ &  $   -3.0046$ &         ---   & \, & $   11.8624$ &         ---   & \, &     ---  \\[\rp]
 \hline\hline
\end{tabular}}
\label{table:h_L}
\end{table}

\section{\label{sec:analytical_tools}Analytical tools}

This section summarizes essential tools used in the present analysis.

\subsection{\label{subsec:isotropy}Isotropy of $\mathcal{L}$ and $\mathcal{J}$}
The isotropy of the operators $\mathcal{L}$ and $\mathcal{J}$ plays the key role to derive a general expression for $h_L$, since it enables us to find similarity solutions.
We recall that an operator $\Op(F)$, where $F$ is any function (or any pair of functions) of $\vec{\zeta}$, is called isotropic if $\Op(F(l_{\ij}\zeta_j))(\bm{\zeta}) = \Op(F(\vec{\zeta}))(l_{\ij}\zeta_j)$ holds for any orthogonal transformation matrices $l_{\ij}$ ($l_{\ij} l_{k\!j} = \delta_{ik}$) \cite{Sone07}.
The $\mathcal{L}$ and $\mathcal{J}$ satisfy this condition and thus are isotropic.

Let us introduce the notation $\zeta_a = \zeta_i a_i$ and $\bar{\zeta}_i = \zeta_i - \zeta_a a_i$ with $a_i$ (or $\vec{a}$) being any fixed unit vector. 
Note that $\bar{\zeta}_i$ (or $\bar{\vec{\zeta}}$) is the projection of $\vec{\zeta}$ onto a plane orthogonal to $\vec{a}$.
Taking a function $f$ of the form $ f(\zeta_a, |\bar{\vec{\zeta}}|)$, which is rotationally invariant about $\vec{a}$, we consider the tensor fields defined by $\mathcal{L}(f)$, $\mathcal{L}(\bar{\zeta}_i f)$, and $\mathcal{L}(\bar{\zeta}_i \bar{\zeta}_j f)$.
Then, thanks to the isotropy of $\mathcal{L}$, they have the following representations \cite{Sone07}:
\begin{align} \label{e:tensor_fields_L}
& 
\left\{
\begin{array}{ll}
     & \mathcal{L}(f) = \mathcal{L}_0(f),
    \quad
    \mathcal{L}(\bar{\zeta}_i f) = \bar{\zeta}_i \mathcal{L}_1(f), \\[1mm]
     & \mathcal{L}(\bar{\zeta}_i \bar{\zeta}_j f) = \bar{\zeta}_i \bar{\zeta}_j \mathcal{L}_2(f) + (\delta_{\ij} - a_ia_j) \mathcal{L}_3(f), 
\end{array}
\right.
\end{align}
where $\mathcal{L}_i(f)$'s on the right-hand sides are functions of $\zeta_a$ and $|\bar{\vec{\zeta}}|$, i.e., $\mathcal{L}_i(f((\zeta_a,|\bar{\vec{\zeta}}|))) = \mathcal{L}_i(f)(\zeta_a,|\bar{\vec{\zeta}}|)$. 
Similarly, considering the tensor fields $\mathcal{J}(f,g)$, $\mathcal{J}(f,\bar{\zeta}_i g)$,
and $\mathcal{J}(\bar{\zeta}_i f,\bar{\zeta}_j g)$, where $f$ and $g$ are functions of $\zeta_a$ and $|\bar{\vec{\zeta}}|$, we have the following representations \cite{Sone07,taguchi_tsuji_2022_jfm}:
\begin{align} \label{e:tensor_fields_J}
\left\{
\begin{array}{ll}
     &  \mathcal{J}(f,g) = \mathcal{J}_0(f,g),
    \quad 
    \mathcal{J}(f,\bar{\zeta}_i g) = \mathcal{J}(\bar{\zeta}_i g,f) = \bar{\zeta}_i \mathcal{J}_1(f,g), \\[1mm]
     & \mathcal{J}(\bar{\zeta}_i f,\bar{\zeta}_j g) = \bar{\zeta}_i \bar{\zeta}_j \mathcal{J}_2(f,g)
 + (\delta_{\ij} - a_ia_j) \mathcal{J}_3(f,g) + \varepsilon_{\ij m} a_m \zeta_a \mathcal{J}_4(f,g).
\end{array}
\right.
\end{align}
Note that $\mathcal{J}_i(f,g)$'s on the right-hand sides are functions of $\zeta_a$ and $|\bar{\vec{\zeta}}|$. Moreover, 
$\mathcal{J}_i(f,g) = \mathcal{J}_i(g,f)$ ($i=0,2,3$) and $\mathcal{J}_4(f,g) = -\mathcal{J}_4(g,f)$.

For example, from \eqref{e:tensor_fields_L} we have $\mathcal{L}(\zeta_\theta f(\zeta_r,|\bar{\vec{\zeta}|}))/\zeta_\theta = \mathcal{L}(\zeta_\varphi f(\zeta_r,|\bar{\vec{\zeta}}|))/\zeta_\varphi = \mathcal{L}_1(f)(\zeta_r,|\bar{\vec{\zeta}}|)$.

\subsection{\label{subsec:symmetry_relation}Symmetry relation}

We have computed $h_L$ in the present study based on the formula \eqref{e:cross_coupling_formula_hL}. This formula relies on the symmetry relation associated with the linearized Boltzmann equation derived in \cite{Takata09-1}, which we briefly explain here.

Suppose a function $\phi=\phi(\bm{x},\bm{\zeta})$ satisfies, in the region outside the unit sphere $|\vec{x}|=1$, the (steady) linearized Boltzmann equation
\begin{align} \label{e:LB}
    \vec{\zeta} \cdot \nabla_x \phi = \frac{1}{k} \mathcal{L}(\phi) + I, \quad r>1,
\end{align}
where the source term $I=I(\bm{x},\bm{\zeta})$ is a given function of $\vec{x}$ and $\vec{\zeta}$.
We also assume that $\phi$ satisfies the following boundary condition on the sphere (i.e., the linearized version of the Maxwell boundary condition for a steady state):
\begin{align} \label{e:BC2}
    & \phi(\vec{x},\vec{\zeta}) = \alpha g_{\rmw}
+ \alpha \mathcal{K}(\phi-g_{\rmw}) + (1-\alpha) \, \phi(\vec{x},\vec{\zeta} - 2 \zeta_r \vec{n}), \quad \zeta_r > 0,
\quad r = 1,
\end{align}
with
\begin{align}
& g_{\rmw}=2 \vec{\zeta} \cdot \vec{c}_{\rmw} + \left(|\vec{\zeta}|^2 - \tfrac{5}{2}\right) d_{\rmw},
\end{align}
where the vector $\vec{c}_{\rmw} = \vec{c}_{\rmw}(\vec{x})$ and the scalar $d_{\rmw}=d_{\rmw}(\bm{x})$ are given functions of $\bm{x}$ ($|\vec{x}|=1$), independent of $\bm{\zeta}$. Since the domain is unbounded, the system should be supplemented by a condition on the asymptotic behavior of $\phi$ at infinity, i.e.,
\begin{align} \label{e:BC1}
    \phi(\bm{x},\bm{\zeta}) \to I_g(\bm{x},\bm{\zeta}) \quad \text{as} \quad r \to \infty,
\end{align}
where $I_g(\vec{x},\vec{\zeta})$ is a given function of $\vec{x}$ and $\vec{\zeta}$.
The system describes the steady behavior of a rarefied gas around a unit sphere in the linearized framework (cf.~\eqref{e:B1}--\eqref{e:inf1}). We have included the additional source term $I$ for later convenience; we will identify it with a source term appearing in the second-order problem (cf. \eqref{e:LB_inner_order2}--\eqref{e:inner_order2_MC})\footnote{Precisely speaking, the form of $g_{\rm w}$ is not general enough to describe $I_{\rm w}^{(2)}$ in \eqref{e:inhomogeneous_2_Iw}. But this does not bother us because the term $\alpha (2 \zeta_\varphi^2 - 1) \,S^2 \sin^2 \theta$ on the r.h.s of \eqref{e:inhomogeneous_2_Iw} does not contribute to $h_L$, and therefore can be eliminated it beforehand.}.

We introduce the notation $g^-(\vec{\zeta}) = g(-\vec{\zeta})$ for any function $g$ of $\vec{\zeta}$.
Further, let us introduce the following notation: $\phi^{\rmI}=\phi^{\rmI}(\bm{x},\bm{\zeta})$ and $\phi^{\rmII}=\phi^{\rmII}(\bm{x},\bm{\zeta})$ are two functions such that
\begin{enumerate}
    \item $\phi^{\rmI}$ satisfies \eqref{e:LB}, \eqref{e:BC2}, and \eqref{e:BC1} with $I=I^{\rmI}$, $I_g=I_g^{\rmI}$, and $g_{\rmw} = g_{\rmw}^{\rmI} \equiv 2 \vec{\zeta} \cdot \vec{c}_{\rmw}^{\rmI} + (|\vec{\zeta}|^2 - \frac{5}{2}) d_{\rmw}^{\rmI}$; and
    \item $\phi^{\rmII}$ satisfies \eqref{e:LB}, \eqref{e:BC2}, and \eqref{e:BC1} with $I=I^{\rmII}$, $I_g=I_g^{\rmII}$, and $g_{\rmw}=g_{\rmw}^{\rmII} \equiv 2 \vec{\zeta} \cdot \vec{c}_{\rmw}^{\rmII} + (|\vec{\zeta}|^2 - \frac{5}{2}) d_{\rmw}^{\rmII}$.
\end{enumerate}
Here, $\phi^{\rmI}$ and $\phi^{\rmII}$ share the same $\mathcal{K}$,  $\mathcal{L}$, and $\alpha$.
Then, 
the following symmetry relation holds \cite{Takata09-1}.
\begin{proposition}
If $\phi^{\rmI}$ and $\phi^{\rmII}$ approach $I_g^{\rmI}$ and $I_g^{\rmII}$ sufficiently fast as $r \to \infty$ so that
\begin{align}
\label{e:condition_asymptotic}
    \lim_{r_0 \to \infty} \int_{|\vec{x}|=r_0} \left \langle \zeta_r (\phi^{\rmI -} - I_g^{\rmI -}) (\phi^{\rmII} - I_g^{\rmII}) \right \rangle \DS = 0,
\end{align}
the following identity, symmetric with respect to the interchange of indices $\rmI$ and $\rmII$, holds:
\begin{align} \label{e:symmetry_identity}
    & \int_{|\vec{x}|=1} \left\langle \zeta_r g_{\rmw}^{\rmII -} \phi^\rmI \right \rangle \DS + \lim_{r_0 \to \infty}\int_{|\vec{x}|=r_0} \left \langle \zeta_r I_g^{\rmII -} \phi^{\rmI} \right \rangle \DS
    \nonumber \\
    & \qquad 
    - \lim_{r_0 \to \infty} \frac{1}{2} \int_{|\vec{x}|=r_0} \left \langle \zeta_r I_g^{\rmII -} I_g^{\rmI} \right \rangle \DS 
    - \int_{|\vec{x}|>1} \left \langle I^{\rmII -} \phi^{\rmI} \right \rangle \DV
    \nonumber \\
    & 
    = \int_{|\vec{x}|=1} \left \langle \zeta_r g_{\rmw}^{\rmI -} \phi^\rmII \right \rangle \DS + \lim_{r_0 \to \infty }\int_{|\vec{x}|=r_0} \left \langle \zeta_r I_g^{\rmI -} \phi^{\rmII} \right \rangle \DS
    \nonumber \\
    & \qquad 
    - \lim_{r_0 \to \infty} \frac{1}{2} \int_{|\vec{x}|=r_0} \left \langle \zeta_r I_g^{\rmI -} I_g^{\rmII} \right \rangle \DS - \int_{|\vec{x}|>1} \left \langle I^{\rmI -} \phi^{\rmII} \right \rangle \DV.
\end{align}
\end{proposition}

Note that the symmetry relation is derived in a more general situation in \cite{Takata09-1}, encompassing other boundary conditions and the cases of bounded and unbounded domains. The result shown above is a particular version of the relation in the case of an unbounded domain.

To apply the symmetry relation, we consider the following two problems: (I) problem of a uniform flow past a sphere for $\phi_{\rmU}^{(1)}$ (Sect.~\ref{subsec:phi_1}); 
and (II) the boundary-value problem for $\phi_{\rmUS}^{(2)}$ (Sect.~\ref{subsubsec:similarity_2}).
Further, we alter the condition at infinity for $\phi_{\rmU}^{(1)}$ to be able to obtain a meaningful result. Specifically, we consider the situation in which the flow is parallel to the $x_3$ direction, i.e., the direction of the transverse force, at infinity. Thus, denoting a solution to the modified problem by $\bar{\phi}_{\rmU}^{(1)}$, we arrive at the following choice:
\begin{align}
\left\{
\begin{array}{ll}
\phi^\rmI = \bar{\phi}_{\rmU}^{(1)},
\quad
g_{\rmw}^\rmI = 0,
\quad
I_g^\rmI = 2 \zeta_3,
\quad
I^\rmI = 0, \\[2mm]
\phi^\rmII = \phi_{\rmUS}^{(2)},
\quad
g_{\rmw}^\rmII = 2 \mathcal{K}(\phi_{\rmU}^{(1)}|_{r=1})
\,\zeta_\varphi \,S \sin \theta,
\quad
I_g^\rmII = 0,
\quad
I^\rmII = \dfrac{2}{k} \mathcal{J}(\phi_{\rmU}^{(1)},\phi_{\rmS}^{(1)}).
\end{array}
\right.
\end{align}
Then, the formula \eqref{e:cross_coupling_formula_hL} follows from the identity \eqref{e:symmetry_identity} after carrying out the integrals with respect to $\theta$ and $\varphi$ 
\footnote{$\bar{\phi}_{\rmU}^{(1)}=\bar{\phi}_{\rmU}^{(1)}(r,\theta,\varphi,\zeta_r,\zeta_\theta,\zeta_\varphi)$ is expressed as $\bar{\phi}_{\rmU}^{(1)} = (\sin \theta \sin \varphi)\,  \phi_{\rmU a}^{(1)}
- (\zeta_\theta \cos \theta \sin \varphi + \zeta_\varphi \cos \varphi)\,
\phi_{\rmU b}^{(1)}$ in terms of $\phi_{\rmU  \beta}^{(1)}=\phi_{\rmU \beta}^{(1)}(r,\zeta_r,\zeta)$, $\beta=a,b$.} \cite{taguchi_tsuji_2022_jfm}.

The symmetry relation was initially used to establish cross relations among weakly perturbed systems described by the linearized Boltzmann equation for arbitrary Knudsen number \cite{Takata09-1,Takata_PHF09_thermophoresis}. Here, we have demonstrated that we can even use the symmetry relation to investigate a nonlinear effect in a weakly nonlinear problem. The point is that the expansion of $\phi$ in $\epsilon$ leads to a sequence of linearized problems with or without source terms, enabling us to apply the symmetry relation. In this sense, weak nonlinearity is crucial for the present method to be applicable. 

Finally, the current approach is, at a glance, similar to the one discussed in \cite{Sharipov_PhysRevE_2011}. However, we should not forget the condition on the asymptotic behavior in the far field \eqref{e:condition_asymptotic}, which ensures the validity of the symmetry relation in an unbounded-domain problem. No account of a similar condition is given in \cite{Sharipov_PhysRevE_2011}. There is no such ambiguity in the symmetry relation formulated in \cite{Takata09-1}. In the present problem, the condition \eqref{e:condition_asymptotic} is fulfilled \cite{taguchi_tsuji_2022_jfm}, and consequently, we can use the symmetry relation to derive the alternative formula for $h_L$. 

\section{Concluding remarks and perspectives}

In this paper, we have considered the flow past a rotating sphere based on kinetic theory. This seemingly simple flow received particular interest owing to the dramatic changes in transverse force over low and high Knudsen numbers (or large and small particle sizes). Our previous study, which is based on the BGK model under the diffuse reflection boundary condition, revealed the force's transition in terms of $\kn$. It also determined the precise value of the threshold above which the transverse force changes the sign. The present study extends these results to the case of the Maxwell boundary condition to get insight into the effect of the surface accommodation on the transverse force. This paper presents results for moderate values of accommodation coefficient $\alpha$, i.e.,  $\alpha=0.8$ and 0.5. Our findings are summarized as follows:
\begin{enumerate}
    \item The magnitude of the transverse force becomes smaller with the decrease of $\alpha$;
    \item The threshold for the transverse force is insensitive to the accommodation coefficient for moderate $\alpha$.
\end{enumerate}
We have followed the same approach as in the previous report, and adopting Maxwell's condition is straightforward. As a byproduct, we have obtained new numerical data for $h_D$ and $h_M$, which describe the drag and torque acting on the sphere (see Table~\ref{table:h_L}).

The values of the accommodation coefficient considered in the present paper are limited to $\alpha=0.8$ and 0.5. Therefore, enriching data for different values of $\alpha$ or using another type of boundary conditions will improve the availability of the formula \eqref{e:lift_dimensional}.
Also, we can apply the method developed in this (and previous) study to investigate different types of nonlinear effects. For example, we can explore the impact of sphere heating (or cooling) on the drag through coupling between the uniform flow and heat transfer (due to the sphere heating/cooling).

\begin{acknowledgement}
The present work was supported by JSPS KAKENHI Grant Nos.~JP20H02067 and JP22K03924.
\end{acknowledgement}


\newcommand{\noop}[1]{}\hyphenation{Post-Script Sprin-ger}

\end{document}